# Compositional Tuning in $Na_xAlB_{14}$ via Diffusion Control

Mihiro Hoshino[1], Suguru Iwasaki[2], Shigeto Hirai[3], Yoshihiko Ihara[4], Tohru Sugahara[5], Haruhiko Morito[6], Masaya Fujioka[1, 7]*

[1] Research Institute for Electronic Science, Hokkaido University, Sapporo 001-0020, Japan

[2] Department of Industrial Chemistry, Faculty of Engineering, Tokyo University of Science, Tokyo 125-8585, Japan

[3] Faculty of Engineering, Kitami Institute of Technology, Kitami 090–8507, Japan

[4] Department of Physics, Faculty of Science, Hokkaido University, Sapporo 060-0810, Japan

[5] Faculty of Materials Science and Engineering, Kyoto Institute of Technology, Kyoto 606－8585, Japan

[6] Institute for Materials Research, Tohoku University, Sendai 980–8577, Japan

[7] Multi-Materials Research Institute, National Institute of Advanced Industrial Science and Technology (AIST), Nagoya 463-8560, Japan

**Abstract**

A uniform Na distribution in $Na_xAlB_{14}$ was achieved using high-pressure diffusion control (HPDC), which promotes Na deintercalation through enhanced diffusion under high pressure, combined with post-annealing. $Na_xAlB_{14}$ with a non-stoichiometric Na composition is thermodynamically metastable, and conventional solid-state reactions with adjusted starting compositions typically result in the formation of stoichiometric $NaAlB_{14}$ and side products. While HPDC alone typically leads to concentration gradients, intentionally halting the Na removal process before complete extraction, followed by annealing, enabled a uniform composition across the bulk. This allowed structural and electronic properties to be examined over a wide range of Na concentrations. As Na content decreased, electrical conductivity increased, and the optical band gap narrowed. NMR measurements showed an increase in the density of states at the Fermi level, consistent with DFT calculations predicting boron-related in-gap states. Boron vacancies at specific sites were found to generate deep levels near the band gap center, which can explain experimentally observed optical gap reduction. These results demonstrate that diffusion-controlling methods can be effectively applied to synthesize metastable compounds with tunable compositions in covalent frameworks. Furthermore, they provide a foundation for designing functional boride-based materials with adjustable electronic properties by controlling Na extraction and inducing defect formation.

Introduction

Elemental diffusion between solids during thermal treatment is an essential phenomenon that facilitates chemical reactions, and conventional solid-state reactions rely on the mutual diffusion of all constituent elements at elevated temperatures. In contrast, topochemical reactions have attracted attention as methods that enable the introduction, extraction, or substitution of specific elements through mild diffusion processes without significantly modifying the overall crystal structure [1-5]. These reactions provide access to metastable states that are often inaccessible via equilibrium pathways.

Recently, synthetic strategies that aim to control the diffusion of specific elements by introducing artificial chemical potential gradients at solid–solid interfaces have been proposed, and anisotropic diffusion has been demonstrated in the Na–Si clathrate system [6]. In addition, electric fields can serve as effective driving forces for diffusion across bulk solids in contact [7]; however, this approach requires the target materials to be insulators to avoid electrostatic screening. Several diffusion control techniques utilizing these driving forces have also been reported [8].

Among various methods, immersing host materials in solutions containing dissolved guest species has traditionally been one of the most common approaches to synthesizing intercalation compounds [9]. In contrast, Proton-Driven Ion Introduction (PDII) is a solvent-free method that diffuses cations into host materials beyond the solid-solid interfaces [10-12]. In PDII, protons are injected from one side of a solid electrolyte, generating a chemical potential gradient that drives cation diffusion into the opposite side. The absence of solvent allows high-temperature treatment and prevents its co-insertion, enabling structural analysis of the intercalation systems like one-dimensional transition metal trichalcogenides, where keeping crystallinity is challenging during ion introduction [13]. These benefits highlight one of the major advantages of solid-state approaches. Furthermore, such solid processes have also demonstrated the intercalation of divalent ions, including the co-introduction of hydrogen to facilitate Mg intercalation into transition metal dichalcogenide [14], and a stepwise method in which prior Na insertion promotes subsequent Ca intercalation into graphite[15].

Not only low-dimensional materials with van der Waals gaps, as discussed above, but also systems with compositional tunability, featuring robust frameworks and diffusible elements, are targets for this synthesis strategy. A wide range of compounds have already been demonstrated to utilize anisotropic diffusion beyond solid–solid interfaces, including the substitution of Na with H in phosphate glass materials [7, 16, 17], the exchange of Na with K in NASICON-type structures [10], the substitution of oxygen sites with hydride in perovskite-type ionic crystals, and the extraction of Na from covalently bonded frameworks such as clathrates [18] and borides [19]. Moreover, such diffusion-based approaches have also been extended to the development of advanced devices, including superconducting wire [20] and protonic ceramic fuel cell (PCFC) fabrication [21].

However, a major challenge in diffusion-controlling techniques is achieving compositionally uniform materials. This is because elemental migration within solids, driven by chemical potential gradients, inherently leads to the development of concentration gradients during the process. $NaAlB_{14}$ is a compound in which Na ions are incorporated within a robust boron framework [22-24], and its Na content can be controlled via high-pressure diffusion control (HPDC), which utilizes anisotropic diffusion under high pressure[19, 25]. The Na concentration can be modulated from $x = 1$ to nearly $x \approx 0$ in $Na_xAlB_{14}$ without altering the boron framework. However, due to the inevitable formation of concentration gradients during diffusion, achieving compositionally uniform intermediate Na concentrations ($0 < x < 1$) throughout the entire bulk material has remained a significant challenge.

This study aims to expand the applicability of diffusion-controlling techniques for achieving compositionally homogeneous bulk materials by intentionally halting the process before complete Na removal, followed by post-annealing to average out the concentration gradient. As a result, homogeneous bulk samples with $x = 0.89$ and $x = 0.44$ were successfully synthesized. This enabled the evaluation of physical properties and crystal structures in the previously unexplored intermediate composition range.

### Sample preparation and structural characterization

$NaAlB_{14}$ powders were synthesized by sintering under Na vapor, as previously reported [23]. The resulting powder was then subjected to high-pressure (HP) annealing at 900 °C for 2 hours under 4 GPa to promote strong interparticle bonding. In addition, applying pressure during HPDC compresses the material in response to volume contraction by Na extraction, thereby suppressing mechanical degradation and facilitating smooth Na diffusion. Together, these high-pressure treatments are essential for enabling controlled compositional tuning through diffusion [26]. The cell assembly used for HPDC is illustrated in Figure 1(a), and the specific roles of each component have been described in detail in previous work [19].

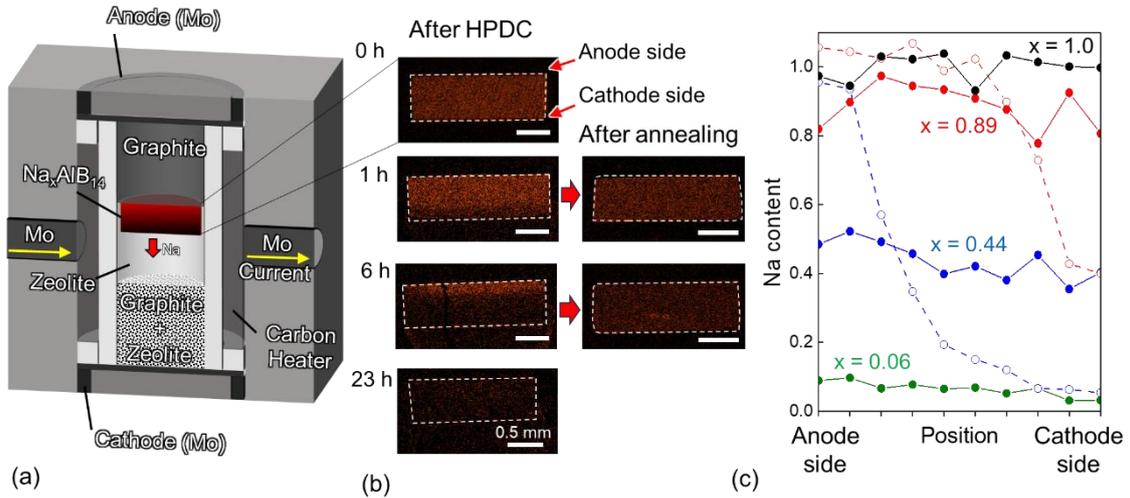

Fig. 1. Schematic of high-pressure diffusion control (HPDC) and evolution of Na distribution. (a) Cell assembly. This panel illustrates the configuration used for HPDC. (b) Energy-dispersive X-ray Spectroscopy (EDS) mapping of Na concentration during HPDC and after post-annealing. Na distribution is visualized across the sample cross-section at different treatment durations, revealing progressive Na depletion from the cathode side and homogenization after annealing. (c) Change in EDS line scan before and after post-annealing. Na migrates from high- to low-concentration regions during HPDC, driven by an internal chemical potential gradient. Post-annealing at 800 °C for 48 h in a vacuum leads to a uniform Na distribution.

In this study, four samples were prepared: one as-HP-annealed sample and three HPDC-treated samples with varying treatment durations. The time-dependent HPDC treatment conditions for each sample are presented in Figure S1(a–c). During HPDC processing, the applied voltage was adjusted to maintain a constant current of 0.2 mA.

Energy-dispersive X-ray Spectroscopy (EDS) mapping images for Na shown in Figure 1(b) reveal the evolution of Na distribution across the sample cross-section during the HPDC treatment, as observed from specimens extracted at different time intervals. As the treatment time increases, the Na concentration gradually decreases from the cathode side. This behavior is further supported by the EDS line analysis presented in Figure 1(c). The observed Na distribution suggests that Na migration is primarily driven by a chemical potential gradient. If the migration were dominated by an externally applied electric field, the Na concentration would be expected to decrease from the anode side [17]. The Na concentration gradient after HPDC was averaged by post-annealing at

800 °C for 48 hours in an evacuated quartz tube, as demonstrated in Figure 1(b) and 1(c). This behavior is attributed to Na diffusion driven by the internal concentration gradient, proceeding from regions of higher to lower chemical potential. As a result, the average Na contents were estimated to be x = 0.89, 0.44, and 0.06, based on the quantitative data shown in Figure 1(c). These values correlate well with the total charge transferred during the HPDC treatment, as shown in Figure S1(a–c), and a linear relationship between charge and Na content is confirmed in Figure S2.

XRD profiles for samples with different Na concentrations are shown in Figure 2(a). Since $NaAlB_{14}$ is harder than tungsten carbide and cannot be pulverized into fine powder, the measurements were performed on the bulk surfaces taken from the anode side after post-annealing. The vertical bars at the bottom of Figure 2(a) indicate the reference diffraction positions of $NaAlB_{14}$. No obvious impurity phases were observed in any of the samples. The 101 peak is highly sensitive to Na content, with its intensity increasing as the Na concentration decreases. In addition, the lattice parameters gradually decrease with decreasing Na content, as shown in Figure 2(b). This trend is reasonable, as the extraction of Na ions leads to a reduction in lattice volume. XRD profiles from both the anode and cathode sides before and after post-annealing are presented in Figures S3(a)-S3(e).

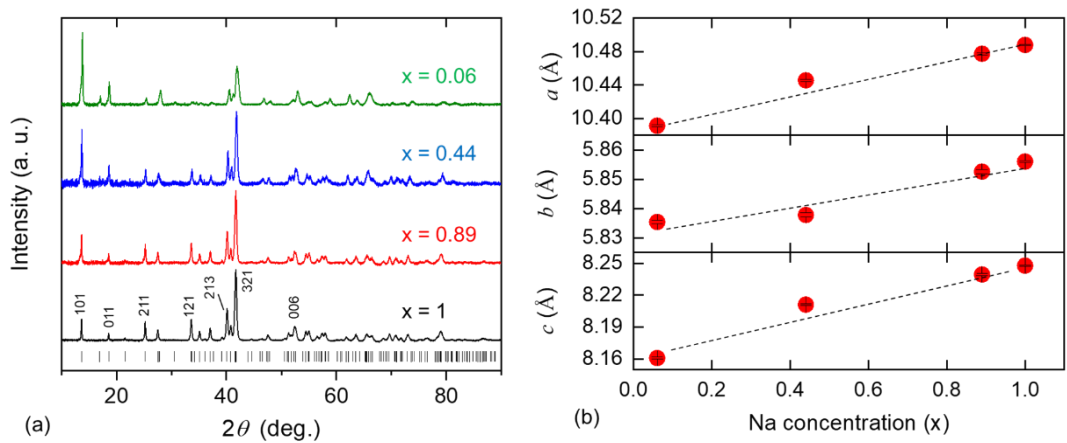

Figure 2. **Structural characterization of Na-extracted $Na_xAlB_{14}$ samples.** (a) XRD patterns obtained from the bulk surface of samples with varying Na content. Vertical bars indicate the reference peak positions of $NaAlB_{14}$. (b) The change in lattice parameters with decreasing Na concentration.

### Electronic properties

Figure 3(a) shows the temperature dependence of resistivity for samples with different

Na concentrations. The as-prepared NaAlB$_{14}$ samples exhibit variations in electrical conductivity, which appear to result from subtle differences in synthesis conditions, possibly due to B-site defects or excess interstitial boron. To minimize the influence of such variations, resistivity measurements were performed on Na-extracted samples obtained from the same batch of NaAlB$_{14}$.

The resistivity at room temperature decreases markedly with decreasing Na content. While the sample with x = 1 exhibits an extremely high resistivity exceeding $10^6$ Ω·cm at 20 °C, the sample with x = 0.06 shows a much lower resistivity of 55 Ω·cm at 20 °C and 1.6 Ω·cm at 500 °C. Although Na extraction is generally expected to induce hole doping, boride compounds are known to exhibit a self-compensation effect, in which changes in carrier concentration are offset by a reconstruction of the boron framework, often involving interstitial boron or boron vacancies [27-29]. As shown in Figure S4, the positive Seebeck coefficient confirms that this material exhibits p-type conduction. This indicates that hole carriers are indeed introduced beyond the level that would be compensated by the self-compensation effect. On the other hand, as shown in Figure S5, the activation energy ($E_a$) estimated from the Arrhenius plots shows a systematic decrease with decreasing Na content: $E_a$ = 604 meV for x = 1, 387 meV for x = 0.89, 147 meV for x = 0.40, and 127 meV for x=0.06. In parallel, the optically estimated band gaps ($E_g$) from diffuse reflectance spectroscopy also decrease with decreasing Na content, from $E_g$ = 2.29 eV for x = 1, 2.11 eV for x = 0.89, 1.18 eV for x = 0.44, to 1.08 eV for x = 0.06 as shown in Figure 3(b). The large difference between $E_a$ and $E_g$ suggests that hole transport with increasing Na extraction arises from hopping via shallow acceptor levels. This behavior is reminiscent of n-type ZnO, which exhibits $E_a$ = 61 meV at 300 K and $E_g$ = 3.37 eV [30]. Notably, the optical band gap exhibits an abrupt reduction from approximately 2 eV at x = 0.89 to nearly 1 eV at x = 0.44, suggesting that Na removal triggers the formation of deep defect levels within the band gap. This behavior may be associated with the creation of boron-related defects or reconstruction of the boron framework, both of which can substantially alter the electronic structure and introduce in-gap states, as discussed in the following section.

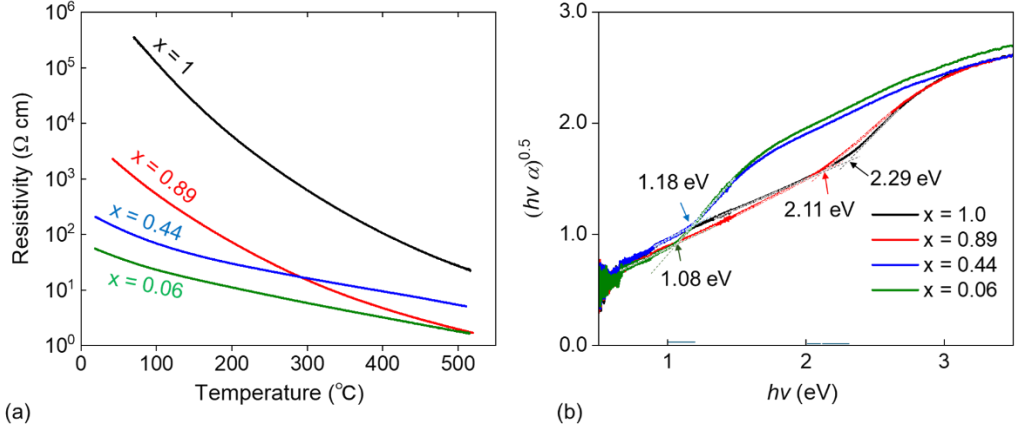

Figure 3. Electronic properties of $Na_xAlB_{14}$. (a) The temperature dependence of resistivity for samples with different Na contents. Resistivity decreases significantly with decreasing Na content. (b) Optical band gaps estimated from diffuse reflectance spectra for samples with different Na contents.

## Discussion

To further elucidate the origin of the enhanced electrical conductivity upon Na extraction, the density of states (DOS) at the Fermi level, $N(E_F)$, was examined by nuclear magnetic resonance (NMR) spectroscopy. In addition, density functional theory (DFT) calculations were used to investigate the formation of in-gap states associated with boron-related defects.

Figure 4(a) presents the estimated $N(E_F)$ based on the nuclear spin-lattice relaxation rate ($1/T_1$) of $^{11}B$ and $^{27}Al$ nuclei, measured at room temperature, using the relation $N(E_F) \propto (1/T_1)^{0.5}$ [31]. The vertical axis is normalized to the value at x ≈ 0. NMR results for each nucleus at different Na concentrations are summarized in Figure S6. A clear increase in carrier density with Na deficiency is directly observed. DFT calculations reveal that the boron-derived states dominate near the Fermi level and the enhancement of boron DOS is consistent with the experimental trend toward increased carrier density and reduced activation energy, as shown in Figure 2. These results highlight the critical role of the boron framework in governing the electronic transport properties while retaining semiconducting behavior.

The crystal structure of $AlB_{14}$ is shown in Figure 4(b) using Visualisation for Electronic and STructural Analysis (VESTA) [32]. One unit cell contains four Al atoms and fifty-six B atoms, corresponding to the chemical formula $Al_4B_{56}$. The $B_{12}$ icosahedral cluster consists of B2, B3, B4, and B5 sites, while the B1 site serves as a bridging unit between neighboring clusters. Note that, for clarity in our DFT analysis, the five B sites in $Al_4B_{56}$ were redefined as B1–B5 based on their local environments. These definitions differ from previous research[19, 22]. Figure 4(c) presents the calculated density of states (DOS) for $Na_4Al_4B_{56}$ and $Al_4B_{56}$ without boron vacancies, as well as for $Al_4B_{55}$ models in which a single vacancy is introduced at each B site. While $NaAlB_{14}$ is a semiconductor with a band gap of approximately 2.2 eV, $Al_4B_{56}$ without boron vacancies is predicted to show metallic behavior due to heavy hole doping. Introducing boron vacancies acts as electron donation, compensating the holes and shifting the Fermi level toward the band edge. The observed semiconducting behavior in Figure 3(a) suggests that such vacancies may play a self-compensating role.

Moreover, calculations indicate that boron vacancies create various in-gap states, including shallow levels that may account for the reduced activation energy observed in Figure 3(a). While the precise identification of boron defect positions requires further investigation, such as synchrotron X-ray or neutron diffraction studies, previous works have shown that the $B_{12}$ icosahedral clusters are structurally very stable[33, 34]. Considering this, the B1 site, which serves as a bridging position between these clusters, is the most probable location for defect formation from a structural standpoint[33, 34]. When a vacancy is introduced at the B1 site, a localized deep-level DOS appears near the center of the band gap, as shown in Figure 4(c). If the concentration of such defects increases with Na extraction, as hypothesized, this may explain the optical band gap narrowing from ~2.2 eV to ~1 eV observed in Figure 3(b). These results suggest that both shallow states, enhancing electrical conductivity, and deep states, reducing the optical band gap, originate from boron vacancy formation and the associated framework reconstruction.

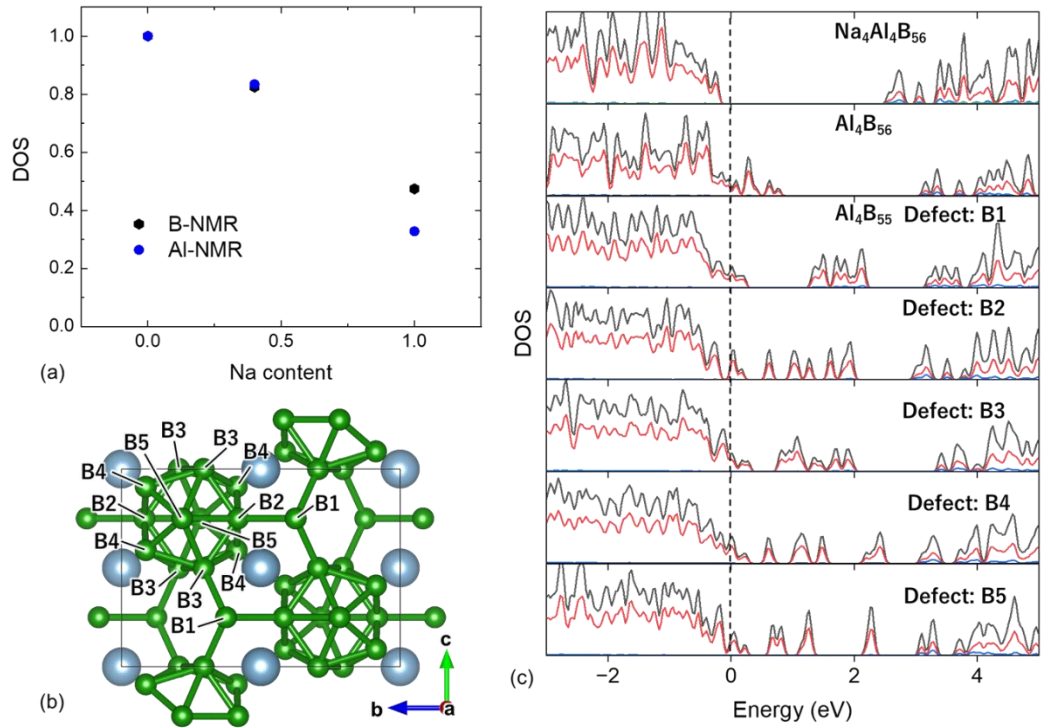

Figure 4. Nuclear Magnetic Resonance (NMR) spectroscopy and Density Functional Theory (DFT) analysis of electronic structure. (a) Density of states at the Fermi level estimated from NMR measurements. (b) Crystal structure of $AlB_{14}$. Five distinct B sites are indicated. The $B_{12}$ cluster is composed of B2, B3, B4, and B5 sites, while the B1 site serves as a bridge between clusters. (c) The DFT-calculated density of states for $Na_4Al_4B_{56}$, $Al_4B_{56}$, and B-deficient $Al_4B_{55}$.

## Conclusion

In this study, homogeneous tuning of Na content in $Na_xAlB_{14}$ was achieved by combining HPDC with post-annealing. Because HPDC relies on diffusion driven by a chemical potential gradient arising from concentration differences, achieving uniform intermediate compositions is inherently challenging. By halting the Na removal process at an intermediate stage and subsequently annealing the sample, uniform Na distribution was successfully obtained. This process enabled the synthesis of thermodynamically metastable Na-extracted compounds with non-stoichiometric compositions, which are typically inaccessible through conventional solid-state reactions. This approach enabled a systematic investigation of the structural and electronic properties in the intermediate composition range (0 < x < 1), which had previously been difficult to access due to compositional inhomogeneity.

The results demonstrate that decreasing Na content leads to a marked increase in electrical conductivity and a narrowing of the optical band gap. These changes are attributed to the formation of boron-related defects, which introduce both shallow and deep in-gap states. NMR measurements revealed an increase in the density of states at the Fermi level with decreasing Na content, consistent with DFT calculations indicating that boron-derived defect states dominate the electronic structure. In particular, boron vacancies at specific sites (B1) were found to introduce deep states near the band gap center, providing a plausible explanation for the observed reduction in the optical band gap.

Overall, this study demonstrates that metastable compounds with intermediate compositions can be accessed by skillfully utilizing diffusion processes. The insights obtained here provide a valuable foundation for designing functional boride-based materials with tunable electronic properties through controlled Na extraction and defect formation.

## Method

Na contents were estimated by SEM−EDS measurements using a JCM-6000 (JEOL). XRD measurements using a MiniFlex 600 with D/teX Ultra (Rigaku) were performed to determine the crystal structures. The temperature dependencies of the resistivities were measured using a handmade machine with the DC four-probe technique. Diffuse reflection measurements were performed using a spectrophotometer, V770 (Jasco). NMR measurements were performed at a fixed magnetic field of 13 T using a phase-coherent NMR spectrometer [35]. The frequency spectrum was obtained by the Fourier transformation of spin echo signal. The relaxation rate was measured by the saturation-recovery method. DFT calculations were performed using the projector augmented-wave (PAW) method as implemented in the Vienna ab initio simulation package (VASP) [36]. The exchange-correlation functional was treated using the Perdew–Burke–Ernzerhof (PBE) formulation within the generalized gradient approximation (GGA) [37] for structural relaxations and initial electronic structure calculations. For improved accuracy in the density of states and defect state analyses, the Heyd–Scuseria–Ernzerhof (HSE06) hybrid functional was additionally employed[38, 39]. The plane-wave cutoff energy 450 eV was used. The Brillouin zone was sampled using a 2 × 4 × 3 Monkhorst-pack grid, and Gaussian smearing with σ = 0.05 eV was applied.

## Acknowledgment

This work was supported by the Japan Science and Technology Agency (JST) CREST (Grant No. JPMJCR19J1), the Japan Society for the Promotion of Science (JSPS) (Grant Nos. 19H02420, 22H04458, 24K06950 and 24K01171). We would like to express our gratitude to Ms. Masae Sawamoto, Research Institute for Electronic Science, Hokkaido University, Ms. Yukino Nishikubo, Innovative Functional Materials Research Institute, National Institute of Advanced Industrial Science and Technology, and Ms. Maki Tsurumoto, Kyoto Institute Technology, for preparing the HPDC cells, measuring Seebeck properties and providing insight that greatly assisted the research.